\documentclass[12pt]{article}
\usepackage{amsmath}
\usepackage{amssymb}
\usepackage{array}
\usepackage{geometry}
\usepackage{graphicx}
\usepackage{enumerate}
\numberwithin{equation}{section}

\begin{document}
\numberwithin{equation}{section}
\begin{center}
{\bf \large EXTENSION OF THERMONUCLEAR FUNCTIONS\\
THROUGH THE PATHWAY MODEL INCLUDING\\
MAXWELL-BOLTZMANN AND TSALLIS DISTRIBUTIONS}\\
\bigskip
{\bf H.J. Haubold}\\
\small{Office for Outer Space Affairs, United Nations, \\
Vienna International Centre, P.O. Box 500, A-1400 Vienna, Austria\\
and}\\
{\bf D. Kumar}\\
{\small
Centre for Mathematical Sciences Pala Campus\\ 
Arunapuram P.O., Palai, Kerala  686 574, India}
\end{center}
{\small{\bf Abstract.} The Maxwell-Boltzmannian approach to nuclear reaction rate theory is extended to cover Tsallis statistics (Tsallis, 1988) and more general cases of distribution functions. An analytical study of respective thermonuclear functions is being conducted with the help of statistical techniques. The pathway model, recently introduced by Mathai (2005), is utilized for thermonuclear functions and closed-form representations are obtained in terms of H-functions and G-functions. Maxwell-Boltzmannian thermonuclear functions become particular cases of the extended thermonuclear functions. A brief review on the development of the theory of analytic representations of nuclear reaction rates is given.\\

\noindent
\begin{center}
{\section{\bf Introduction}}
\end{center}

In the evolution of the Universe, chemical elements are created in cosmological and stellar nucleosynthesis (Clayton, 1983; Fowler, 1984). Solar nuclear energy generation and solar neutrino emission are governed by chains of nuclear reactions in the gravitationally stabilized solar fusion reactor (Davis, 2003). One of the first utilization of Gamow's quantum mechanical theory of potential barrier penetration to other than the analysis of radioactive decay was to the question on how do stars generate energy (Critchfield, 1972; for a brief essay on the history of this discovery see Mathai and Haubold, 1988). Continued attempts are aiming at generating energy through controlled thermonuclear fusion in the laboratory. In a nuclear plasma, the rate of reactions and thus energy release can be determined by an average of the Gamow penetration factor over the distribution of velocities of the particles of the plasma (Haubold and Mathai, 1984). Understanding the mathematical and statistical methods for the evaluation of thermonuclear reaction rates is one of the goals of research in the field of stellar and cosmological nucleosynthesis. Practically all applications of fusion plasmas are controlled in some way or another by the theory of thermonuclear reaction rates under specific circumstances. After several decades of effort, a systematic and complete theory of thermonuclear reaction rates has been developed (Haubold and John, 1978; Anderson et al., 1994; Haubold and Mathai, 1998; Mathai and Haubold, 2002). Reactions between individual particles produce a randomization of the energy and velocity distributions of particles. The depletion of particles by reactions is balanced by the diffusion of the particles in the macroscopically inhomogeneous medium. As a result of this balance, the fusion plasma may reach a quasi-stationary state close to equilibrium, in which steady fluxes of matter, energy, and momentum are present. This also lead to the assumption that the distribution of particles can be assumed to be Maxwell-Boltzmannian in almost all cases of interest to stellar physics and cosmology.\\

The derivation of closed-form representations of nuclear reaction rates and useful approximations of them are based on statistical distribution theory and the theory of generalized special functions, mainly in the categories of Meijer's G-function and Fox's H-function (Mathai and Saxena, 1973; Mathai, 1993; Aslam Chaudhry and Zubair, 2002). For an overview on the application and historical background of integrals and distribution functions for reaction rates and their representation in terms of special functions, see Hegyi (1999) and Moll (2002). Special cases which can be derived from the general theory through expansion of respective physical parameters (like the cross section factor) are nonresonant (Hussein and Pato, 1997; Ueda et al., 2000) and resonant (Ueda et al., 2004; Newton et al., 2007) reaction rates, reaction rates for cosmological nucleosynthesis (Bergstroem et al., 1999) and fitting of experimental data to analytic representations (Brown and Jarmie, 1990). Specific mathematical methods for deriving approximate analytic representations of nuclear reaction rates are expansions of hypergeometric functions (Saigo and Saxena, 1998), transformation of extended gamma functions (Aslam Chaudhry, 1999), and asymptotic expansion of the Laplace transform of functions (Ferreira and Lopez, 2004).\\

Only recently, related to the production of neutrinos in the gravitationally stabilized solar fusion reactor, the question of possible deviations of the velocity distribution of plasma particles from the Maxwell-Boltzmannian case due to memory effects and long-range forces has been raised (Coraddu et al., 1999; Lavagno and Quarati, 2002; Coraddu et al., 2003; Lissia and Quarati, 2005; Lavagno and Quarati, 2006). This was initiated by Tsallis' non-additive generalization of Boltzmann-Gibbs statistical mechanics which generates q-exponential function as the fundamental distribution function instead of the Maxwell-Boltzmann distribution function. Tsallis statistics covers Boltzmann-Gibbs statistics for the case q = 1 (Tsallis, 1988; Gell-Mann and Tsallis, 2004; Tsallis, 2004). A first attempt was made to extend the theory of nuclear reaction rates from Maxwell-Boltzmann theory to Tsallis theory (Saxena et al., 2004; Mathai, 2005; Mathai and Haubold, 2007). This paper develops the complete theory for closed-form representations of nuclear reaction rates for Tsallis statistics. In this context an interesting discovery was made by Mathai (2005), namely that even more general distribution functions can be incorporated in the theory of nuclear reaction rates by appealing to entropic and distributional pathways. Starting from a generalized entropy of order $\alpha$, through the maximum entropy principle, distribution functions are generated which include Maxwell-Boltzmann and Tsallis distributions as special cases.\\

In Subsections 1.1, 1.2, 1.3, and 1.4, we introduce the definition of the thermonuclear reaction rate and respective thermonuclear functions for the standard, cut-off, depleted, and screened case, respectively. Each subsection provides the integral of the thermonuclear function for the cases of Maxwell-Boltzmann distribution and $\alpha$ distribution, the latter covers the q-exponential of Tsallis. Section 2 provides prerequisites for the use of G- and H- functions in Mellin-Barnes integral representation and also discusses briefly the pathway model of Mathai. Section 3 elaborates the closed-form representation of the thermonuclear functions for the case of Maxwell-Boltzmann and $\alpha$ distributions in terms of G- and H-functions. Section 4 provides conclusions. \\

From Mathai and Haubold (1988) it can be seen that the expression for the reaction rate $r_{ij}$ of the reacting particles $i$ and $j$
taking place in a nondegenerate environment is
\begin{eqnarray}
r_{ij} &=& n_i n_j \left( \frac{8}{\pi \mu}\right)^{\frac{1}{2}}
{\left(\frac{1}{kT}\right)}^{\frac{3}{2}} \int _0^{\infty} E
\sigma(E) {\rm e}^{-\frac{E}{k T}} {\rm d} E \\
&=& n_i n_j \langle {\sigma {v}} \rangle \nonumber
\end{eqnarray}
where $n_i$ and $n_j$ are the number densities of the particles $i$
and $j$, the reduced mass of the particles is denoted by $\mu=
\frac{m_i m_j}{m_i+m_j},~T $  is the temperature, $k$ is the
Boltzmann constant, the reaction cross section is $\sigma(E)$ and
the kinetic energy of the particles in the center of mass system is
$E= \frac{\mu{v}^2}{2} $ where $v$ is the relative velocity of
the
interacting particles $i$ and $j$.\\

The reaction probability is written in the form $\langle {\sigma {v}
} \rangle$ to indicate that it is an appropriate average of the
product of the reaction cross section and relative velocity of the
interacting particles.  For detailed physical interpretations see
Haubold and Mathai (1984).\\

{\subsection{\bf Standard non-resonant thermonuclear function}}

For non-resonant nuclear reactions between two nuclei of
charges $z_i$ and $z_j$ colliding at low energies below the Coulomb
barrier, the reaction cross section has the form (Mathai and Haubold, 2002)
\begin{eqnarray*}
\sigma(E)&=& \frac{S(E)}{E}  {\rm e}^{-2 \pi \eta(E)}\\
\noalign{with} \eta(E)&=&\left(\frac{\mu}{2}\right)^{\frac{1}{2}}
\frac{z_i z_j {\rm e}^2}{\hbar E^{\frac{1}{2}}}
\end{eqnarray*}
where $\eta(E) $ is the Sommerfeld parameter, $\hbar$ is Planck's quantum of action, ${\rm e}$ is the quantum of electric
charge, the cross section factor $S(E)$ is often found to be a
constant or a slowly varying function of energy over a limited range
of energy (Mathai and Haubold, 1988). The cross section factor
$S(E)$ may be expressed in terms of the power series expansion,
\[ S(E)= S(0) + \frac{{\rm d}S(0)}{{\rm d}E}E +\frac{1}{2}
\frac{{\rm d}^2S(0)}{{\rm d}E^2}E^2. \]
Then
\begin{eqnarray}
\langle {\sigma
\nu } \rangle  &=& \left( \frac{8}{\pi
\mu}\right)^{\frac{1}{2}}\sum_{\nu=0}^2{\left(\frac{1}{kT}\right)}^
{-\nu+\frac{1}{2}}\frac{S^{(\nu)}(0)}{\nu !}\times
\int_0^{\infty}E^\nu {\rm e}^{-\frac{E}{kT}-2 \pi \eta(E)}{\rm d}E \nonumber\\
&=&\left( \frac{8}{\pi\mu}\right)^{\frac{1}{2}}
\sum_{\nu=0}^2{\left(\frac{1}{kT}\right)}^{-\nu+\frac{1}{2}}
\frac{S^{(\nu)}(0)}{\nu !}\times
\int_0^{\infty}x^\nu {\rm e}^{-x-bx^{-\frac{1}{2}}}{\rm d}x
\end{eqnarray}
where $x=\frac{E}{kT}$ and $b=\left(\frac{\mu}{2kT}\right)^{\frac{1}{2}}
\frac{z_i z_j {\rm e}^2}{\hbar }.$ \\

The collision probability integral, called thermonuclear function, for non-resonant thermonuclear
reactions in the Maxwell-Boltzmannian case is (Haubold and Mathai,
1984)
\begin{equation}
I_1( \nu ,1 , b , \frac{1}{2} )=\int_0^{\infty}x^\nu {\rm
e}^{-x-bx^{-\frac{1}{2}}}{\rm d}x.
\end{equation}
We will consider here the general integral
\begin{equation}
I_1(\gamma-1,a,b,\rho)= \int_0^{\infty}x^{\gamma-1} {\rm
e}^{-ax-bx^{-\rho}}{\rm d}x,~~ a>0, b>0, \gamma>0, \rho>0.
\end{equation}\\

{\subsection{\bf Non-resonant thermonuclear function with high energy cut-off}}

Usually, the thermonuclear fusion plasma is assumed to be in thermodynamic
equilibrium.  But if there appears a cut-off of the high energy tail
of the Maxwell-Boltzmann distribution function in $(1.3)$ then the
thermonuclear function is given by
\begin{equation}
I_2^{(d)}( \nu ,1 , b , \frac{1}{2} )=\int_0^d x^\nu {\rm
e}^{-x-bx^{-\frac{1}{2}}}{\rm d}x,~~ b>0, \gamma>0, \nu>0.
\end{equation}
Again we consider the general form of the integral (1.5) as
\begin{equation}
I_2^{(d)}(\gamma-1,a,b,\rho)=\int_0^dx^{\gamma-1} {\rm
e}^{-ax-bx^{-\rho}}{\rm d}x,~~ a>0, b>0, \gamma>0, \rho>0.
\end{equation}
For physical reasons for the cut-off modification of the
Maxwell-Boltzmann distribution function of the relative kinetic
energy of the reacting particles refer to the paper Haubold and Mathai (1984).\\

{\subsection{\bf Non-resonant thermonuclear function with depleted tail}}

A depletion of the high energy tail of the Maxwell-Boltzmann
distribution function of the relative kinetic energies of the nuclei
in the fusion plasma is discussed in Haubold and Mathai (1984).\\

For the thermonuclear function, in comparison
to the strict Maxwell-Boltzmannian case, we have the integral
\begin{equation}
I_3(\nu,1,1,\delta,b,\frac{1}{2}) =\int_0^{\infty}x^{\nu} {\rm
e}^{-x-x^{\delta} -bx^{-\frac{1}{2}}}{\rm d}x,~~b>0, \gamma>0.
\end{equation}
We will consider the general integral of the type
\begin{equation}
I_3(\gamma-1, a , z , \delta , b ,\rho)= \int_0^{\infty}x^{\gamma-1}
{\rm e}^{-ax-zx^\delta-bx^{-\rho}}{\rm d}x,
\end{equation}
where  $z>0,a>0, b>0, \gamma>0, \rho>0.$\\

{\subsection{\bf Non-resonant thermonuclear function with screening}}

The plasma correction to the fusion process due to a static or
dynamic potential, ie the electron screening effects for the
reacting particles, the collision probability integral to be
evaluated in the case of  the screened non-resonant nuclear reaction
rate is (see Haubold and Mathai, 1984)
\begin{equation}
I_4(\nu,1 , b, t,\rho)=\int_0^{\infty}x^\nu {\rm
e}^{-x-b(x+t)^{\frac{1}{2}}}{\rm d}x.
\end{equation}
In this case we consider the general integral as
\begin{equation}
I_4(\gamma-1, a , b, t,\rho)=\int_0^{\infty}x^{\gamma-1} {\rm
e}^{-ax-b(x+t)^{\frac{1}{2}}}{\rm d}x,~~ t>0, \rho>0, b>0,a>0
\end{equation}
where $t$ is the electron screening parameter.\\

In the following we are evaluating  the thermonuclear reaction
probability integrals by a using pathway model (Mathai, 2005). In Section 2 we give
the basic definitions that we require in this paper. We evaluate the
integrals $I_1$ and  $I_2$  by implementing Mathai's pathway model
and represent each of them in terms of H-function and
G-function in Section 3.\\

\begin{center}
{\section{\bf Mathematical Preliminaries}}
\end{center}

We need some basic quantities for our discussion, which will be
defined here. The gamma function denoted by $ \Gamma (z) $ for
complex number $z$ is defined as \begin {equation}
\Gamma (z)= \int_0^{\infty} t^{z-1} {\rm e}^{-t} {\rm d} t, \Re (z)>0
\end{equation}
where $\Re (\cdot)$ denotes the real part of $(\cdot)$. In general $\Gamma
(z)$ exists for all values of $z$, positive or negative, except at
the points $z= 0,-1,-2,\cdots $. These are the poles of $\Gamma
(z)$. But the integral representation holds for the real part of $z$
to be positive. Another important result we need is the
multiplication formula.
 If $z$ is any complex number, $z\neq0,-1,-2,...$ and let $m $ be a positive integer then the
multiplication formula for gamma functions is
\begin{equation}
\Gamma(mz)= {(2\pi)}^\frac{1-m}{2} m^{mz-{\frac{1}{2}}} \Gamma(z)
\Gamma \left( z+\frac{1}{m}\right)\cdots \Gamma \left(
z+\frac{m-1}{m}\right).
\end{equation}
For $m=2$ we get the
duplication formula for gamma functions,
\begin{equation}
\Gamma(2z)= {\pi}^{-\frac{1}{2}} 2^{2z-1}\Gamma(z) \Gamma \left(
z+\frac{1}{2}\right).
\end{equation}

The Mellin transform of a real scalar function $f(x)$ with parameter
$s$ is defined as
\begin{equation}
M_f(s)=\int_0^{\infty} x^{s-1} f(x) {\rm d}x
\end{equation}
whenever $M_f(s)$ exists.
If $f_1(x)$ and $f_2(y)$ are integrable functions  on the positive
reals and if $x^k f_1(x)$ and $y^kf_2(y) $ are absolutely
integrable, then the Mellin convolution property is defined as
\begin{equation}
f_3(u)=\int_0^{\infty} \frac{1}{x} f_1(x) f_2(\frac{u}{x}) {\rm
d}x,
\end{equation}
then the Mellin transform of $ f_3 $ denoted by $ M_{f_3}(s) $  is

\begin{equation}
M_{f_3}(s)=M_{f_1}(s)M_{f_2}(s)
\end{equation}
where $M_{f_1}(s)=\int_0^{\infty} x^{s-1} f_1(x) {\rm d}x$ and
$M_{f_2}(s)=\int_0^{\infty} y^{s-1} f_2(y) {\rm d}y.$ \\

\noindent
The type-1 beta integral is defined as
\begin{equation}
\int_0^1 x^{\alpha-1} (1-x)^{\beta-1} {\rm d} x= \int_0^1
y^{\beta-1} (1-y)^{\alpha-1} {\rm d} y=\frac{\Gamma(\alpha)
\Gamma(\beta)}{\Gamma(\alpha+\beta)},
\end{equation}
$$\Re(\alpha)>0, \Re(\beta)>0.$$

\noindent
The type-2 beta integral is defined as
\begin{equation}
\int_0^{\infty} x^{\alpha-1}(1+x)^{-(\alpha+\beta)}{\rm d}x=
\int_0^{\infty} y^{\beta-1}(1+y)^{-(\alpha+\beta)}{\rm
d}x=\frac{\Gamma(\alpha) \Gamma(\beta)}{\Gamma(\alpha+\beta)},
\end{equation}
$$\Re(\alpha)>0, \Re(\beta)>0.$$

The G-function which is originally due to C. S. Meijer in 1936 (See
Mathai, 1993; Mathai and Saxena, 1973) is defined as a Mellin-Barnes type integral as follows.
\begin{equation}
G_{p,q}^{m,n}\left(z \big|^{ a_1,\cdots,a_p}_{b_1,\cdots,b_q}\right)
=\frac{1}{2 \pi i}\int_L
g(s) z^{-s}{\rm d}s
\end{equation}
where $i=\sqrt{-1}$, $L$ is a suitable contour and $z\neq 0$,
$m,n,p,q$ are integers,  $0\leq m \leq q$ and $0\leq n\leq p$,
\begin{equation}
g(s)=\frac{\left\{\prod_{j=1} ^{m}\Gamma(b_j+s)\right\} \left\{ \prod_{j=1}
^{n}\Gamma(1-a_j-s)\right\}}{\left\{ \prod_{j=m+1} ^{q}\Gamma(1-b_j-s)\right\} \left\{
\prod_{j=n+1} ^{p}\Gamma(a_j+s)\right\}}
\end{equation}
the empty product is interpreted as unity and the parameters $a_1,a_2,\cdots,a_p$ and
$b_1,b_2,\cdots,b_q$ are complex numbers such that no poles of
$\Gamma(b_j+s),~j=1,\cdots,m$ coincides with any pole of
$\Gamma(1-a_k-s),~k=1,\cdots,n;  $
\[-b_j-\nu \neq 1-a_k+\lambda,~~j=1,\cdots,m ;~~ k=1,\cdots,n;~~\nu,\lambda=0,1,\cdots. \]

This means that $a_k-b_j\neq\nu+\lambda+1$ or $a_k-b_j$ is not a
positive integer for$~~j=1,\cdots,m ;~ k=1,\cdots,n$.  We also
require that there is a strip in the complex $s$-plane which
separates the poles of $\Gamma(b_j+s),~ j=1,\cdots,m$ from those of
$\Gamma(1-a_k-s),~k=1,\cdots,n  $ (See Mathai (1993) for the
conditions of validity).  The H-function is defined as
\begin{eqnarray}
&&H_{p,q}^{m,n}\left(z \big|^
{(a_1,{\alpha}_1),\cdots,(a_p,{\alpha}_p)}_
{(b_1,{\beta}_1),\cdots,(b_q,{\beta}_q)}  \right)=
H(z) \nonumber\\
&=&\frac{1}{2 \pi i}\int_C \frac{\left\{ \prod_{j=1} ^{m}\Gamma(b_j+\beta_j s)\right\}
\left\{\prod_{j=1}
^{n}\Gamma(1-a_j- \alpha_j s)\right\} }{ \left\{ \prod_{j=m+1}
^{q}\Gamma(1-b_j-\beta_j s)\right\} \left\{ \prod_{j=n+1}
^{p}\Gamma(a_j+\alpha_j s)\right\}} z^{-s}{\rm d}s
\end{eqnarray}
which is a Mellin-Barnes type integral where $0\leq m \leq q,~0\leq
n \leq p,~\alpha_j~(j=1,2,\cdots,p)$,~ $\beta_j ~(j=1,2,\cdots,q)$ are positive real numbers and $a_j
~(j=1,2,\cdots,p)$ and $b_j~(j=1,2,\cdots,q)$ are complex numbers.  Any empty product is considered as unity and it is assumed that no poles of $\Gamma(b_j+\beta_js)$ for $j=1,2,\cdots,m $
coincides with any pole of $\Gamma(1-a_j- \alpha_j s)$ for
$j=1,2,\cdots,n$. Furthermore, $C$ is a contour in the complex
$s$-plane moving from $c-i\infty$ to $c+i\infty$ for some real
number $c$ such that the points $s=\frac{b_j+\nu }{\beta_j}$ for
$j=1,2,\cdots,m $ and $\nu =0,1,\cdots$ and the points
$s=\frac{1-a_k+\lambda }{\alpha_k}$ for $k=1,2,\cdots,n$ and $\lambda =0,1,\cdots
$ lie to the left and right of $C$, respectively. For details see
Springer (1979) and Mathai (1993).\\

Next we utilize the pathway model of Mathai (2005). When fitting a
model to experimental data very often one needs a model for a distribution function from a given
parametric family, or sometimes we may have a situation of the right
tail cut-off.  In order to take care of these situations and going
from one functional form to another, a pathway parameter $\alpha$ is
introduced, see Mathai (2005) and Mathai and Haubold (2007).  By
this model we can proceed from a generalized type-1 beta model to a
generalized type-2 beta model to a generalized gamma model when the
variable is restricted to be positive. For the real scalar case the
pathway model is the following,
\begin{equation}
f(x)=c|x|^{\gamma-1}[1-a(1-\alpha)|x|^\delta]^{\frac{1}{1-\alpha}},~~
a>0,\delta>0, 1-a(1-\alpha)|x|^\delta>0, \gamma>0,
\end{equation}
where $c$ is the
normalizing constant and $\alpha$ is the pathway parameter. When
$\alpha<1$ the model becomes a generalized type-1 beta model in the
real case.  This is a model with the right tail cut-off. When
$\alpha>1$ we have $1-\alpha=-(\alpha-1), \alpha>1$ so that
\begin{equation}
f(x)=c|x|^{\gamma-1}[1+a(\alpha-1)|x|^\delta]^{-\frac{1}{\alpha-1}},
\end{equation}
which is a generalized type-2 beta model for real $x$. When
$\alpha\rightarrow1$ the above two forms will reduce to
\begin{equation}
f(x)=c|x|^{\gamma-1}{\rm e}^{-ax^\delta}.
\end{equation}
Observe that the normalizing constant $c$ appearing in (2.12),
(2.13) and (2.14), are different.\\

\begin{center}
{\section {\bf Evaluation of the Integrals of Thermonuclear Functions}}\end{center}

Now we will evaluate the  integrals $I_1$and  $I_2$
by introducing the pathway model.\\

{\subsection{\bf Evaluation of $I_1$ }}

\[ I_1= \int _{ 0 }^{\infty} x^{\gamma -1}{\rm e}^{-ax-bx^{-{\rho}}}
{\rm d}x ,~ a>0, b>0,
{\gamma }>0, {\rho }>0. \]

Replace  ${\rm e}^{-ax} $ by $[1+a(\alpha-1 )x]^{-\frac{1}{\alpha-1
}} $. As $ \alpha \rightarrow 1, $ $[1+a(\alpha-1
)x]^{-\frac{1}{\alpha-1 }} $ becomes $ {\rm e}^{-ax} $ so that we
can extend the integral $I_1$ to a wide class through the  pathway
parameter $\alpha$.  Let us denote the wider class of this integral
by $I_{1\alpha}$.
\begin{equation*}
I_{1\alpha}=\int _{ 0 }^{\infty} x^{\gamma -1}[1+a(\alpha-1 )x]^{-\frac{1}
{\alpha-1 }}{\rm e}^{-bx^{-\rho }}{\rm d}x
\end{equation*}
where$ ~\alpha >1,~a>0,~\delta =1,~1+a(\alpha-1 )x>0,~\rho
>0,~b>0. $
This is the product of two integrable functions.
Hence we can apply Mellin convolution property for finding the value of the integral.
Here let us take
\begin{eqnarray}
f_1(x) &=& \left \{ \begin {array}{ll}  x^{\gamma} [1+a(\alpha-1 )x]^{-\frac{1}{
\alpha-1 }} & \mbox{for } 0\leq x<\infty , a>0, \gamma >0, \alpha >1\\
0 , &  \mbox{otherwise. }\\
\end {array}\right.\\
f_2(y) &=& \left \{ \begin {array}{ll}   {\rm e}^{-y^\rho } & \mbox{for }
0\leq y<\infty , ~~~ \rho >0\\
0 ,&  \mbox{otherwise. }\\
\end {array}\right.
\end{eqnarray}
From (2.5) we have,
\begin{eqnarray}
I_{1\alpha } &=& \int _{ 0 }^{\infty} \frac{1}{x} f_1(x) f_2 (\frac{u}{x}) {\rm d}x \nonumber \\
&=& \int _{ 0 }^{\infty}x^{\gamma-1} [1+a(\alpha-1 )x]^{-\frac{1}{(\alpha-1 ) }}
{\rm e}^{-u^\rho x^{-\rho }}{\rm d}x\nonumber \\
&=& \int _{ 0 }^{\infty}x^{\gamma-1} [1+a(\alpha-1 )x]^{-\frac{1}{(\alpha-1) }}
{\rm e}^{-bx^{-\rho }}{\rm d}x~~ \mbox{where} ~~b=u^\rho, \alpha >1.
\end {eqnarray}
The Mellin transform of $f_3=I_{1\alpha }$ is the product of the Mellin
transforms of $f_1$ and $f_2$
\begin{eqnarray*}
M_{f_3}(s)&=& M_{f_1}(s)M_{f_2}(s) \\
M_{f_1}(s)&=& \int _{0}^{\infty}x^{s-1}x^{\gamma}
{[1+a(\alpha-1 )x]}^{-\frac{1}{(\alpha-1 ) }}{\rm d}x \\
&=& \int _{0}^{\infty}x^{\gamma + s-1}{[1+a(\alpha-1 )x]}^{-\frac{1}{(\alpha-1 )}}
{\rm d}x.
\end{eqnarray*}
Putting $a(\alpha-1 )x=t$, we get
\begin{eqnarray}
M_{f_1}(s)&=&\frac{1}{[a(\alpha-1 )]^{\gamma +s}}\int _ 0 ^ \infty
t^{\gamma +s-1}{(1+t)}^{-\frac{1} {\alpha-1}}{\rm d}t \nonumber \\
&=&\frac{1}{[a(\alpha-1 )]^{\gamma +s}}\frac{\Gamma (\gamma +s)
\Gamma \left ( \frac{1}{\alpha-1 }-\gamma-s\right )}{\Gamma \left
(\frac{1}{\alpha-1 }  \right )},
\end {eqnarray}
where $~~ \Re(\gamma +s)>0,~~\Re( \frac{1}{\alpha-1 }-\gamma-s)>0,~\alpha >1.$
\begin{equation*}
M_{f_2}(s)= \int _{ 0 }^{\infty}y^{s-1}{\rm e}^{-y^\rho } {\rm d}y.
\end{equation*}
Putting $y^{\rho} = t $,we get
\begin{eqnarray}
M_{f_2}(s)&=& \frac{1}{\rho } \int _{ 0 }^{\infty} t^{\frac{s}{\rho }-1}
{\rm e}^{-t} {\rm d}t \nonumber \\
&=&\frac{1}{\rho } \Gamma {\left ( \frac{s}{\rho } \right
)},~~\Re(s)>0.
\end{eqnarray}
From $(3.4)$ and $(3.5)$ \\
\begin {equation*}
M_{f_3}(s)=\frac{1}{\rho [a(\alpha-1 )]^{\gamma +s}}\frac{\Gamma
(\gamma +s)\Gamma \left ( \frac{1}{\alpha-1 }-\gamma-s \right )
\Gamma {\left ( \frac{s}{\rho } \right )}}{\Gamma \left
(\frac{1}{\alpha-1} \right )}.
\end{equation*}
Then the density of $u $ denoted by  $ f_3(u) $ is available from
the inverse Mellin transform.\\
\begin{multline}
f_3(u) = \frac{1}{\rho [a(\alpha-1 )]^{\gamma }\Gamma \left
(\frac{1}{\alpha-1 } \right )} \frac{1}{2 \pi i} \int_{c-i \infty
}^{c+ i \infty } \Gamma (\gamma +s)\Gamma \left ( \frac{1}{\alpha-1
}-\gamma-s \right)  \Gamma {\left ( \frac{s}{\rho } \right )}\\
\times [a(\alpha-1 )u]^{-s} {\rm d} s.
\end {multline}
Comparing (3.3) and (3.6)
\begin {multline}
\int _{ 0 }^{\infty} x^{\gamma -1}[1+a(\alpha-1 )x]^{-\frac{1}{\alpha-1 }}
{\rm e}^{-bx^{-\rho }}{\rm d}x \\
=\frac{1}{\rho [a(\alpha-1 )]^{\gamma }\Gamma \left
(\frac{1}{\alpha-1 } \right )} H_{1,2}^{2,1} \left(a(\alpha-1
)b^{\frac{1}{\rho }}\big| ^ {\left (1- \frac{1}{\alpha-1 }+\gamma ,
1 \right) }_
{(\gamma ,1)(0,\frac{1}{\rho })} \right),~ \alpha >1. \nonumber \\
\end {multline}
Therefore,
\begin{equation*}
I_{1\alpha} =\frac{1}{\rho [a(\alpha-1 )]^{\gamma }\Gamma \left
(\frac{1}{\alpha-1 } \right )} H_{1,2}^{2,1} \left(a(\alpha-1
)b^{\frac{1}{\rho }}\big|^ {\left (1- \frac{1}{\alpha-1 }+\gamma , 1
\right) }_ {(\gamma ,1)(0,\frac{1}{\rho })} \right)
\end{equation*}
where $a>0,~b>0,~ \gamma>0,~\rho>0,\alpha>1,  ~\Re(s)>0,~
\Re(\gamma+s)>0$. Note that when $\alpha\rightarrow1,~I_{1\alpha}$
becomes $I_{1}$. But $ I_{1\alpha} $ contains all neighborhood
solutions for various values of
$\alpha$  for $\alpha>1$.\\

\noindent
{\bf Special Case:}
\medskip

If $\frac{1}{\rho}$ is an integer then put $ \frac{1}{\rho}= m $.
Using equation (2.2) we get,
\begin{equation*}
I_{1\alpha}
=\frac{\sqrt{m}(2 \pi)^{\frac{1-m}{2}}}{ [a(\alpha-1 )]^{\gamma }\Gamma \left
(\frac{1}{\alpha-1 } \right )}G_{1,m+1}^{m+1,1}
\left(\frac{a(\alpha-1)b^m}{m^m} \big|^{1-\frac{1}{\alpha-1 }+\gamma
}_{ 0,\frac{1}{m},\cdots,\frac{m-1}{m},\gamma} \right)
\end{equation*}
where $a>0,~b>0,~ \gamma>0,~\alpha>1 $.\\

In the thermonuclear function for a non-resonant thermonuclear reaction in the
Maxwell-Boltzmannian case $\gamma-1=\nu,~a=1,\rho =\frac{1}{2}$,
then by using (2.3) we get,
\begin{equation*}
I_{1\alpha}=\frac{1}{[(\alpha-1 )]^{\gamma }\Gamma \left
(\frac{1}{\alpha-1 } \right )}G_{1,3}^{3,1} \left(
\frac{(\alpha-1)b^2}{4} \big|^{1-\frac{1}{\alpha-1}+\nu+1}_
{0, \frac{1}{2}, \nu+1} \right)
\end{equation*}
where $b>0,~ \nu>0,\alpha>1 ~$.\\

{\subsection{\bf Evaluation of $I_2$}}

Replace  ${\rm e}^{-ax} $ by $[1-a(1-\alpha )x]^{\frac{1}{1-\alpha }}
$. As $ \alpha \rightarrow 1, ~[1-a(1-\alpha
)x]^{\frac{1}{1-\alpha }}$ becomes $ {\rm e}^{-ax} $. Let us denote
the two integrals by $I_{2}^{(d)}$ and $I_{2\alpha}^{(d)}$
respectively.
\begin{eqnarray*}
{I_2^{(d)}}&=&\int _{ 0 }^{d} x^{\gamma -1}{\rm e}^{-ax-bx^{-{\rho}}}
{\rm d}x ,~~ a>0, b>0, {\gamma }>0, {\rho }>0. \\
I_{2\alpha}^{(d)}&=&\int _{ 0 }^{d} x^{\gamma -1}[1-a(1-\alpha )x]^{\frac{1}
{1-\alpha }}{\rm e}^{-bx^{-\rho }}{\rm d}x
\end{eqnarray*}
where $ d\leq\frac{1}{a(1-\alpha )},~\alpha <1,~a>0,~\delta
=1,~1-a(1-\alpha )x>0,~\rho >0,~b>0$.  For convenience of
integration let us assume that $d=\frac{1}{a(1-\alpha)}$.  Then
$I_{2\alpha}^{(d)}$ is the product of two integrable functions.
Hence we can apply Mellin convolution property for finding the value of the integral.
Here let us take
\begin{eqnarray}
f_1(x) &=& \left \{ \begin {array}{ll}  x^{\gamma}
[1-a(1-\alpha )x]^{\frac{1}{1-\alpha }} & \mbox{for } 0\leq x<{\frac{1}
{a(1-\alpha )}} , a>0, \gamma >0, \alpha <1\\
0 ,&  \mbox{otherwise. }\\
\end {array}\right.\\
f_2(y) &=& \left \{ \begin {array}{ll}   {\rm e}^{-y^\rho } & \mbox{for }
0\leq y<\infty ,~ \rho >0\\
0 ,&  \mbox{otherwise. }\\
\end {array}\right.
\end{eqnarray}
From (2.5) we have,
\begin{eqnarray}
I_{2\alpha }^{(d)}&=& \int _{ 0 }^{\infty} \frac{1}{x} f_1(x) f_2 (\frac{u}{x}) {\rm d}x \nonumber \\
&=& \int _{ 0 }^{d}x^{\gamma-1} [1-a(1-\alpha )x]^{\frac{1}{(1-\alpha ) }}
{\rm e}^{-u^\rho x^{-\rho }}{\rm d}x ~~\mbox{where} ~ d=\frac{1}{a(1-\alpha )}\nonumber \\
&=& \int _{ 0 }^{d}x^{\gamma-1} [1-a(1-\alpha )x]^{\frac{1}{(1-\alpha) }}
{\rm e}^{-bx^{-\rho }}{\rm d}x~~ \mbox{where}
~~b=u^\rho, \alpha <1.
\end {eqnarray}
The Mellin transform of $f_3=I_{2\alpha }^{(d)}$ is the product of the Mellin
transforms of $f_1$ and $f_2$.
\begin{eqnarray*}
M_{f_3}(s)&=& M_{f_1}(s)M_{f_2}(s) \\
M_{f_1}(s)&=& \int _{0}^{\frac{1}{a(1-{\alpha}) }}x^{\gamma + s-1}
{[1-a(1-\alpha )x]}^{\frac{1}{(1-\alpha )}}{\rm d}x
\end{eqnarray*}
Putting $ a(1-\alpha )x = t,$  we get
\begin{eqnarray}
M_{f_1}(s)&=&\frac{1}{[a(1-\alpha )]^{\gamma +s}}\int _ 0 ^ 1 t^{\gamma +s-1}
{(1-t)}^{\frac{1}{1-\alpha }}{\rm d}t \nonumber \\
&=&\frac{1}{[a(1-\alpha )]^{\gamma +s}}\frac{\Gamma (\gamma +s)
\Gamma \left ( \frac{1}{1-\alpha }+1 \right )}{\Gamma \left
(1+\gamma +\frac{1}{1-\alpha }+s  \right )},~~ \Re(\gamma +s)>0, \alpha <1.~~
\end{eqnarray}
From equation $(3.5)$ we get
\begin{equation}
 M_{f_2}(s)=\frac{1}{\rho } \Gamma {\left ( \frac{s}{\rho } \right )},~~~~\Re
(s)>0.
\end{equation}
From $(3.10)$ and $(3.11)$ \\
\begin{equation*}
M_{f_3}(s)=\frac{1}{\rho [a(1-\alpha )]^{\gamma +s}}\frac{\Gamma
(\gamma +s)\Gamma \left ( \frac{1}{1-\alpha }+1 \right ) \Gamma
{\left ( \frac{s}{\rho } \right )}}{\Gamma \left (1+\gamma
+\frac{1}{1-\alpha  }+s \right )}.
\end{equation*}

Then the density of $u $ denoted by  $ f_3(u) $ is available
from the inverse Mellin transform.\\
\begin{equation}
f_3(u)=\frac{\Gamma \left ( \frac{1}{1-\alpha }+1 \right )}{\rho
[a(1-\alpha )]^{\gamma }} \frac{1}{2 \pi i} \int_{c-i \infty }^{c+ i
\infty } \frac{\Gamma (\gamma +s) \Gamma {\left ( \frac{s}{\rho }
\right )}}{\Gamma \left (1+\gamma +\frac{1}{1-\alpha  } +s \right )}
[a(1-\alpha )u]^{-s} {\rm d} s.
\end{equation}
Comparing (3.9) and (3.12)
\begin{eqnarray*}
&&\int _{ 0 }^{d} x^{\gamma -1}[1-a(1-\alpha )x]^{\frac{1}{1-\alpha
}}
{\rm e}^{-bx^{-\rho }}{\rm d}x  \\
&=& \frac{\Gamma \left ( \frac{1}{1-\alpha }+1 \right )} {\rho
[a(1-\alpha )]^{\gamma }} \frac{1}{2 \pi i} \int_{c-i \infty }^{c+ i
\infty } \frac{\Gamma (\gamma +s) \Gamma {\left ( \frac{s}{\rho }
\right )}}{\Gamma \left (1+\gamma +\frac{1}{1-\alpha } +s \right )}
[a(1-\alpha )b^{\frac{1}{\rho }}]^{-s} {\rm d} s  \\
&=&\frac{\Gamma \left ( \frac{1}{1-\alpha }+1 \right )} {\rho
[a(1-\alpha )]^{\gamma }} H_{1,2}^{2,0} \left(a(1-\alpha )
b^{\frac{1}{\rho }}\big|^ {\left ( 1+\gamma  + \frac{1}{1-\alpha } ,
1 \right) }_{ (\gamma ,1)(0,\frac{1}{\rho })} \right), \alpha <1.
\end{eqnarray*}
Therefore,
\begin{equation*}
I_{2\alpha}^{(d)} =\frac{\Gamma \left ( \frac{1}{1-\alpha }+1 \right
)} {\rho [a(1-\alpha )]^{\gamma }} H_{1,2}^{2,0} \left(a(1-\alpha
)b^{\frac{1}{\rho }}\big|^
{\left (1+ \gamma  + \frac{1}{1-\alpha }  , 1 \right) }_
{(\gamma ,1)(0,\frac{1}{\rho })} \right) \nonumber
\end{equation*}
where $a>0, b>0, \gamma>0, \rho>0, \alpha <1, \alpha \rightarrow1$.\\

\noindent
{\bf Special Case:}\\

If $\frac{1}{\rho}$ is an integer then put $ \frac{1}{\rho}= m $.
Then following through the same procedure as before one has
\begin{equation*}
I_{2\alpha}^{(d)}
=\frac{\sqrt{m} {(2 \pi )^{\frac{1-m}{2}}}
\Gamma \left ( \frac{1}{1-\alpha }+1 \right )}{
[a(1-\alpha )]^{\gamma }} G_{1,m+1}^{m+1,0}
\left(\frac{a(1-\alpha)b^m}{m^{m}}
\big|^{1+\gamma +\frac{1}{1-\alpha }}_
{0,\frac{1}{m},\cdots,\frac{m-1}{m},\gamma} \right)
\end{equation*}
where $a>0, b>0,\gamma>0, \alpha <1 $.\\

For the thermonuclear function fornon-resonant thermonuclear reactions with high energy cut
off $\gamma-1=\nu,~a=1,~\rho =\frac{1}{2}$ then we get,
\begin{equation*}
I_{2\alpha}^{(d)}
=\frac{\Gamma \left ( \frac{1}{1-\alpha }+1 \right )}
{\sqrt{\pi} [(1-\alpha )]^{\nu+1 }}
G_{1,3}^{3,0} \left( \frac{(1-\alpha )b^2}{4} \big|^{\nu
+\frac{1}{1-\alpha }+2}_
{0,\frac{1}{2},\nu+1} \right)
\end{equation*}
where $ b>0,\nu>0, \alpha <1 $. The importance of the above result
is that $I_{2\alpha}^{(d)}$ gives an extension of the integral
$I_{2}^{(d)}$ to a wider class of integrals through the pathway
parameter $\alpha$, and their solutions.\\

\begin{center}
{\section{\bf Conclusion}}
\end{center}

In the field of stellar, cosmological, and controlled fusion, for example, the core of the Sun is considered as the gravitationally stabilized solar fusion reactor. The probability for a thermonuclear reaction to occur in the solar fusion plasma depends mainly on two factors. One of them is the velocity distribution of the particles in the plasma and is usually given by the Maxwell-Boltzmann distribution of Boltzmann-Gibbs statistical mechanics. The other factor is the particle reaction cross-section that contains the dominating quantum mechanical tunneling probability through a Coulomb barrier, called Gamow factor. Particle reactions in the hot solar fusion plasma will occur near energies where the product of velocity distribution and tunneling probability is a maximum. The product of velocity distribution function and penetration factor is producing the Gamow peak. Mathematically, the Gamow peak is a thermonuclear function. In case of taking into consideration electron screening of reactions in the hot fusion plasma, the Coulomb potential may change to a Yukawa-like potential. Taking into account correlations and long-range forces in the plasma, the Maxwell-Boltzmann distribution may show deviations covered by the distribution predicted by Tsallis statistics in terms of cut-off or depletion of the high-velocity tail of the distribution function. In this paper, closed-form representations have been derived for thermonuclear functions, thus for the Gamow peak, for Boltzmann-Gibbs and Tsallis statistics. For this purpose, a generalized entropy of order $\alpha$ and the respective distribution function, have been considered. The case $\alpha = 1$ recovers the Maxwell-Boltzmannian case. This general case is characterized by moving cut-off, respectively the upper integration limit of the thermonuclear function to infinity. The closed-form representations of thermonuclear functions are achieved by using generalized hypergeometric functions of Fox and Meijer or H-functions and G-functions, respectively.\\
\begin{center}
{\bf Acknowledgment}
\end{center}

The authors would like to thank the Department of Science and Technology, Government of India, New Delhi, for the financial assistance for this work under project No. SR/S4/MS:287/05, and the Centre for Mathematical Sciences for providing all facilities.\\

\begin{center}
{\bf References}
\end{center}
{\small
\begin{description}
\item
Anderson, W.J., Haubold, H.J., and Mathai, A.M.: 1994, Astrophysical thermonuclear functions, {\it Astrophysics and Space Science}, {\bf 214}, 49-70.
\item
Aslam Chaudhry, M.: 1999, Transformation of the extended gamma function $\Gamma^{2,0}_{0,2}[(B,X)$] with applications to astrophysical thermonuclear functions, {\it Astrophysics and Space Science}, {\bf 262}, 263-270.
\item
Aslam Chaudhry, M. and Zubair, S.M.: 2002, {\it On a Class of Incomplete Gamma Functions with Applications}, Chapman and Hall/CRC, New York.
\item
Bergstroem, L., Iguri, S., and Rubinstein, H.: 1999, Constraints on the variation of the fine structure constant from big bang nucleosynthesis, {\it Physical Review D}, {\bf 60}, 045005-1:9.
\item
Brown, R.E. and Jarmie, N.: 1990, Differential cross sections at low energies for $^2H(d,p)\;^3H\\
\mbox{and}\; ^2H(d,n)\;^3He$, {\it Physical Review C}, {\bf 41}, 1391-1400.
\item
Clayton, D.D.: 1983, {\it Principles of Stellar Evolution and Nucleosynthesis}, Second Edition, The University of Chicago Press, Chicago and London.
\item
Coraddu, M., Kaniadakis, G., Lavagno, A., Lissia, M., Mezzorani, G., and Quarati, P.: 1999, Thermal distributions in stellar plasmas, nuclear reactions and solar neutrinos, {\it Brazilian Journal of Physics}, {\bf 29}, 153-168.
\item
Coraddu, M., Lissia, M., Mezzorani, G., and Quarati, P.: 2003, Super-Kamiokande hep neutrino best fit: a possible signal of non-Maxwellian solar plasma, {\it Physica A}, {\bf 326}, 473-481.
\item
Critchfield, C.L.: 1972, Analytic forms of thermonuclear function, in {\it Cosmology, Fusion and Other Matters: George Gamow Memorial Volume}, Ed. F. Reines, University of Colorado Press, Colorado, pp. 186-191.
\item
Davis Jr., R.: 2003, A half-century with solar neutrinos, {\it Reviews of Modern Physics}, {\bf 75}, 985-994.
\item
Fowler, W.A.: 1984, Experimental and theoretical nuclear astrophysics: the quest for the origin of the elements, {\it Reviews of Modern Physics}, {\bf 56}, 149-179.
\item
Gell-Mann, M. and Tsallis, C. (Eds.): 2004, {\it Nonextensive Entropy: Interdisciplinary Applications}, Oxford University Press, New York.
\item
Haubold, H.J. and John, R.W.: 1978, On the evaluation of an integral connected with the thermonuclear reaction rate in closed-form, {\it Astronomische Nachrichten}, {\bf 299}, 225-232.
\item
Haubold, H.J. and Mathai, A.M.: 1984, On  nuclear reaction rate theory, {\it Annalen der Physik (Leipzig)}, {\bf 41}, 380-396.
\item
Haubold, H.J. and Mathai, A.M.: 1998, An integral arising frequently in astronomy and physics, {\it SIAM Review}, {\bf 40}, 995-997.
\item
Hegyi, S.: 1999, A powerful generalization of the NBD suggested by Peter Carruthers, in {\it Proceedings of the VIII International Workshop on Multiparticle Production}, Eds. T. Csoergo, S. Hegyi, G. Jancso, and R.C. Hwa, Matrahaza, Hungary, World Scientific, Singapore, pp. 272-286.
\item
Hussein, M.S. and Pato, M.P.: 1997, Uniform expansion of the thermonuclear reaction rate formula, {\it Brazilian Journal of Physics}, {\bf 27}, 364-372.
\item
Ferreira, C. and Lopez, J.L.: 2004, Analytic expansions of thermonuclear reaction rates, {\it Journal of Physics A: Mathematical and General}, {\bf 37}, 2637-2659.
\item
Lavagno, A. and Quarati, P.: 2002, Classical and quantum non-extensive statistics effects in nuclear many-body problems, {\it Chaos, Solitons and Fractals}, {\bf 13}, 569-580.
\item
Lavagno, A. and Quarati, P.: 2006, Metastability of electron-nuclear astrophysical plasmas: motivations, signals and conditions, {\it Astrophysics and Space Science}, {\bf 305}, 253-259.
\item
Lissia, M. and Quarati, P.: 2005, Nuclear astrophysical plasmas: ion distribution functions and fusion rates, {\it europhysics news}, {\bf 36}, 211-214.
\item
Mathai, A.M.: 1993, {\it A Handbook of Generalized Special Functions for Statistical and Physical Sciences}, Clarendon Press, Oxford.
\item
Mathai, A.M.: 2005, A Pathway to matrix-variate gamma and normal densities, {\it Linear Algebra and its Applications}, {\bf 396}, 317-328.
\item
Mathai, A.M. and Haubold, H.J.: 1988, {\it Modern Problems in Nuclear and Neutrino Astrophysics}, Akademie-Verlag, Berlin.
\item
Mathai, A.M. and Haubold, H.J.: 2002, Review of mathematical techniques applicable in astrophysical reaction rate theory, {\it Astrophysics and Space Science}, {\bf 282}, 265-280.
\item
Mathai, A.M. and Haubold, H.J.: 2007, Pathway model, superstatistics, Tsallis statistics and a generalized measure of entropy, {\it Physica A} {\bf 375}, 110-122.
\item
Mathai, A.M. and Saxena, R.K.: 1973, {\it Generalized Hypergeometric Functions with Applications in Statistics and Physical Sciences},
Springer-Verlag, Lecture Notes in Mathematics {\bf Vol. 348}, Berlin, Heidelberg, New York.
\item
Moll, V.H.: 2002, The evaluation of integrals: a personal story, {\it Monthly Notices of the American Mathematical Society}, {\bf 49}, 311-317. 
\item
Newton, J.R., Iliadis, C., Champagne, A.E., Coc, A., Parpottas, Y., and Ugalde, C.: 2007, Gamow peak in thermonuclear reactions at high temperatures, {\it Physical Review C}, {\bf 75}, 045801-1:4.
\item
Saigo, M. and Saxena, R.K.: 1998, Expansions of 4F3 when the upper parameters differ by integers, {\it Kyungpook Mathematical Journal}, {\bf 38}, 293-299.
\item
Saxena, R.K., Mathai, A.M., and Haubold, H.J.: 2004, Astrophysical thermonuclear functions for Boltzmann-Gibbs statistics and Tsallis statistics, {\it Physica A}, {\bf 344}, 649-656.
\item
Springer, M.D.: 1979, {\it The Algebra of Random Variables}, Wiley, New York.
\item
Tsallis, C.: 1988, Possible generalization of Boltzmann-Gibbs statistics, {\it Journal of Statistical Physics}, {\bf 52}, 479-487.
\item
Tsallis, C.: 2004, What should a statistical mechanics satisfy to reflect nature?, {\it Physica D}, {\bf 193}, 3-34.
\item
Ueda, M., Sargeant, A.J., Pato, M.P., and Hussein, M.S.: 2000, Effective astrophysical $S$ factor for nonresonant reactions, {\it Physical Review C}, {\bf 61}, 045801-1:6.
\item
Ueda, M., Sargeant, A.J., Pato, M.P., and Hussein, M.S.: 2004, Resonances and thermonuclear reaction rates for charged particle collisions, {\it Physical Review C}, {\bf 70}, 025802-1:6.
\end{description}}
\end{document}